\documentclass[a4paper, amsfonts, amssymb, amsmath, reprint, showkeys, nofootinbib, twoside]{revtex4-1}

\usepackage[paperwidth=210mm,paperheight=297mm,centering,hmargin=1.6cm,vmargin=2.5cm]{geometry}
\usepackage[english]{babel}
\usepackage[utf8]{inputenc}
\usepackage[hang,small,bf]{caption}
\usepackage{booktabs}
\usepackage[colorinlistoftodos, color=green!40, prependcaption]{todonotes}
\usepackage{amsthm}
\usepackage{physics}
\usepackage{xcolor}
\usepackage{graphicx}
\usepackage{mathptmx}
\usepackage{newtxmath}
\usepackage{adjustbox}
\usepackage{placeins}
\usepackage[T1]{fontenc}
\usepackage{lipsum}
\usepackage{csquotes}
\usepackage{float}
\usepackage{caption}
\usepackage{hyperref} 
\usepackage{upgreek}

\newcommand \nm {\ensuremath{\ \mathrm{nm}}}
\newcommand \mm {\ensuremath{\ \mathrm{mm}}}
\newcommand \dB {\ensuremath{\ \mathrm{dB}}}

\begin{document}

%%%%%%%%%%%%%%%%%%%%%%%%%%  Title  %%%%%%%%%%%%%%%%%%%%%%%%%%

\title{High-performance, adiabatically nanotapered fibre-chip couplers in silicon at 2 microns wavelength}

%%%%%%%%%%%%%%%%%%%%%%%%%%  Author list  %%%%%%%%%%%%%%%%%%%%%%%%%%

\author{Dominic A. Sulway$^{1,2, \dagger}$, Yuya Yonezu$^{3,4, \dagger}$, Lawrence M. Rosenfeld$^{2}$, Pisu Jiang$^{2}$, Takao Aoki$^{3}$, Joshua W. Silverstone$^{2,}$}

%%%%%%%%%%%%%%%%%%%%%%%%%%  Affiliations  %%%%%%%%%%%%%%%%%%%%%%%%%%

\email[Correspondence email address: ]{josh.silverstone@bristol.ac.uk}
\affiliation{$^{1}$Quantum Engineering Centre for Doctoral Training, Nanoscience and Quantum Information Building, University of Bristol, Bristol BS8 1FD, UK\\
$^{2}$Quantum Engineering Technology Labs, H. H. Wills Physics Laboratory and Department of Electrical and Electronic Engineering, University of Bristol, Bristol BS8 1FD, UK\\
$^{3}$Department of Applied Physics, Waseda University, Okubo 3-4-1, Shinjuku, Tokyo, Japan\\
$^{4}$NTT Basic Research Laboratories, NTT Corporation, Morinosato-Wakamiya, Atsugi, Kanagawa 243-0198, Japan\\
$^{\dagger}$Authors contributed equally to this work.}

\date{\today}

%%%%%%%%%%%%%%%%%%%%%%%%%%  Abstract  %%%%%%%%%%%%%%%%%%%%%%%%%%

\begin{abstract}
Fibre optic technology connects the world through the Internet, enables remote sensing, and connects disparate functional optical devices. Highly confined silicon photonics promises extreme scale and functional integration. However, the optical modes of silicon nanowire waveguides and optical fibres are very different, making efficient fibre-chip coupling a challenge. Vertical grating couplers, the dominant coupling method today, have limited optical bandwidth and are naturally out-of-plane. Here we demonstrate a new method that is low-loss, broadband, easily manufacturable, and naturally planar. We adiabatically couple a tapering silicon nanowire waveguide to a conic nanotapered optical fibre, measuring transmission between $2.0\ $\textmu m and $2.2\ $\textmu m wavelength. The silicon chip is fabricated at a commercial foundry and then post-processed to release the tapering nanowires. We estimate an optimal per-coupler transmission of $-0.48\dB$ (maximum; 95\% confidence interval [+0.46, --1.68]$\dB$) and a 1-$\dB$ bandwidth of $295\nm$. With automated measurements, we quantify the device tolerance to lateral misalignment, measuring a flat response within $\pm0.968\ $\textmu m. This design can enable low-loss modular systems of integrated photonics irrespective of material and waveband.
\end{abstract}

\maketitle

%%%%%%%%%%%%%%%%%%%%%%%%%%  Body  %%%%%%%%%%%%%%%%%%%%%%%%%%

%%%%%%%%%%%%%%%%%%%%%%%%%%  Introduction  %%%%%%%%%%%%%%%%%%%%%%%%%%

\section{Introduction} \label{sec:intro}

High-performance, optical input/output (IO) is a crucial enabler for applications of integrated classical and quantum photonics. Mid-infrared (MIR) distributed sensing protocols, comprising photonic integrated circuits (PICs) networked by fibre, promise to revolutionise the way we monitor our atmosphere \cite{Jin_2020, Nedeljkovic_2016, Sherif_2016}. The bit-error rates of optical transceivers found in data centres are proportional to the IO transmission of the transceiver \cite{Sugawara_2012}. And low-loss, PIC-to-PIC interfaces between different material platforms may solve some of the issues arising from the stringent requirements on loss, and switching speed, found in quantum photonic information processing schemes \cite{Bourassa_2021}. 

Key metrics for fibre-IO are efficiency, bandwidth, footprint, fabrication process simplicity, ease of alignment and packaging; some metrics are more important than others, depending on the target application. Vertical grating couplers (VGC) can be made reasonably low-loss, small-footprint and facilitate wafer scale testing with V-groove arrays \cite{Li_2013}, but suffer from narrow bandwidths, and high-efficiency devices require complex fabrication processes \cite{Benedikovic_2015}. The multimode, interferometric nature of VGCs means that mode profile and impedance match to a fibre mode will always be imperfect. Inverted taper edge couplers \cite{Almeida_2003} benefit from both efficiency and broadband response \cite{Wang_2016}, but again, mode profile and impedance matching is difficult without specialist lensed or high-numerical aperture fibres. Leading examples of these coupling methods in can be found summarised in Tab. \textbf{\ref{coupler_records}}.

\begin{table}[ht]
    \centering
    \caption{Summary of state-of-the-art silicon photonic fibre input/output device performance. Results shown for CMOS compatible technology.}
    \begin{tabular}[t]{lcccc}
        \hline
        Method &\ $\lambda$\ ($\nm$) &\ $\eta$\ ($\dB$) &\ $\Delta\lambda$\ ($\nm$) &\ Ref\\
        \hline
        Grating coupler & $1535$ & $-1.50$ & 54 (3-dB) & \cite{Li_2013}\\
        Edge coupler & $1550$ & $-1.10$ & 150 (0.3-dB) & \cite{Wang_2016}\\
        Adiabatic coupler & $2003$ & $-0.48$ & $295$ (1-dB) & This work\\
        \hline
        \label{coupler_records}
    \end{tabular}
    \vspace{-5mm}
\end{table}

We propose and demonstrate a coupling regime based on the adiabatic coupling of a tapered optical fibre with a suspended, tapering silicon waveguide, on a post-processed, foundry-fabricated, silicon-on-insulator (SOI) PIC. We demonstrate, high efficiency, wide bandwidth, and robust mechanical alignment. Our demonstration is at MIR wavelengths around $2.1\ $\textmu m, an emerging platform for quantum photonics \cite{Rosenfeld_2020}, though this approach can apply equally to other high-index-contrast photonic platforms at other wavelengths, with minor adjustments.

Adiabatic energy transfer between discrete eigenmodes emerged to describe crossing of molecular energy levels\cite{Zener_1932}, and is now used widely in integrated optics \cite{Shani_1991, Queralto_2018}, tapered optical fibres \cite{Orucevic_2007, Stiebeiner_2010, Aoki_2010, Hoffman_2014, Nagai_2014} and adiabatic evanescent couplers \cite{Groblacher_2013, Tiecke_2015, Patel_2016, Daveau_2017}. Adiabatic energy transfer involves two optical modes, and in our case, these are the isolated fundamental modes of two waveguides---the fibre and the silicon nanowire---with isolated propagation constants $\beta_\mathit{a}$ and $\beta_\mathit{b}$. Such transfer suppresses coupling to higher-order modes by varying either or both of $\beta_\mathit{a}$ or $\beta_\mathit{b}$ slowly along the axis of propagation, such that $\beta_\mathit{b}$ crosses $\beta_\mathit{a}$. The more similar $\beta_\mathit{a}$ and $\beta_\mathit{b}$, the slower the variation must be to maintain adiabatic transfer.

\begin{figure*}[ht!]
	\centering
	\includegraphics[width = 1
	\linewidth]{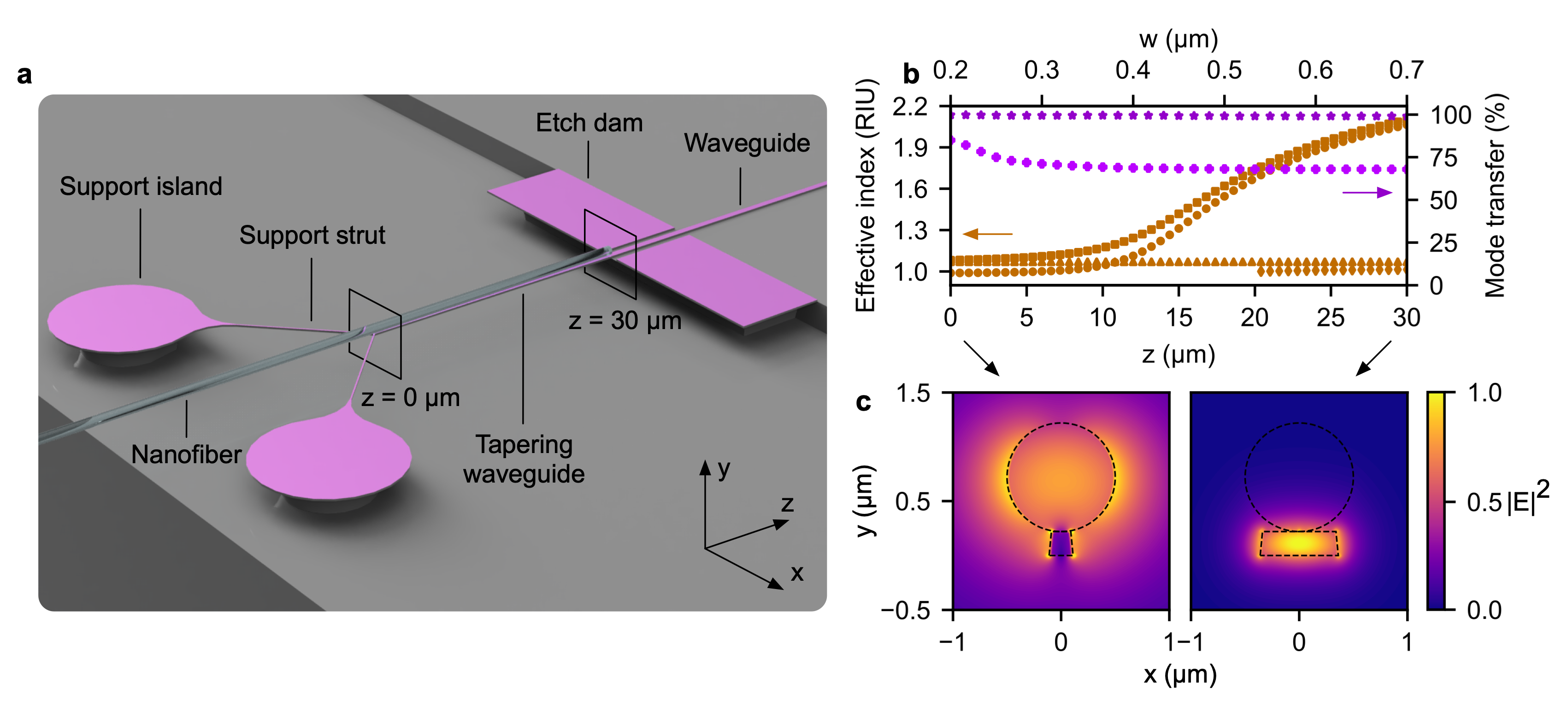}
    \caption{Nanotapered fibre coupler (NTFC) concept and device simulation. \textbf{a} Rendered schematic of the coupled system. The input nanofibre tip (left, transparent) is coupled to a nanotapered, suspended silicon nanowire (purple) with selectively etched buried-oxide (BOX, grey). \textbf{b} Finite-difference time-domain (FDTD, Lumerical) simulation of the $\mathrm{TE_{0}}$ coupling between the constant radii nanotapered fibre, and the nanotapered, suspended silicon nanowire (stars). $\mathrm{TE_{0}}$ coupling with the strut-suspended structure (plusses). Finite-difference eigenmode (FDE, Lumerical) analysis of the effective index evolution of the coupled structure (squares, even supermode), on-chip tapering nanowire (circles, $\mathrm{TE_{0}}$), constant radii nanotapered fibre (triangles, $\mathrm{LP_{01}}$), and coupled structure (diamonds, odd supermode), as a function of $z$ and waveguide width ($w$). \textbf{c} Mode field profiles of the coupled NTFC at $z=0\ $\textmu m and $z=30\ $\textmu m.}
	\label{Fig.1:}
\end{figure*}

%%%%%%%%%%%%%%%%%%%%%%% Design and simulation %%%%%%%%%%%%%%%%%%%%%%%%%

\section{Design and Simulation} \label{sec:theory}

We bring a nanofibre into contact with a suspended, tapering silicon waveguide, as shown in Fig. \textbf{\ref{Fig.1:}}\textbf{a}. Coupling between the fundamental $\mathrm{LP_{01}}$ mode of the fibre and the $\mathrm{TE_{0}}$ mode of the nanowire can be achieved adiabatically. We used a mode solver (Lumerical MODE) to simulate the modes of the isolated and coupled waveguides to initially assess feasibility with the foundry-specified bottom oxide (BOX) thickness $t_\mathit{BOX}$, silicon device layer thickness $t_\mathit{si}$, and the nanofibre diameter $d_\mathit{f}$, assuming complete removal of the top oxide (TOX) layer. With this knowledge, the silicon nanowire start and termination widths, ($w_\mathit{s}$, $w_\mathit{t}$), and the taper length, $L_\mathit{c}$, can be determined. The adiabatic coupling can be characterised by the adiabaticity criteria\cite{Snyder_2012}: $L_\mathit{c}>L_\mathit{b}$, where $L_\mathit{b}=\lambda/(n_\mathit{eff_\mathit{e}}-n_\mathit{eff_\mathit{o}})$ is the beat length; here $\lambda$ is the free-space wavelength, and $n_\mathit{eff_\mathit{e}}$ and $n_\mathit{eff_\mathit{o}}$ are the effective indices of the even and odd supermodes, respectively.

For our case of $\lambda=2.071\ $\textmu m, $t_\mathit{BOX}=3.0\ $\textmu m, $t_\mathit{si}=220\nm$, and $d_\mathit{f}=1.00\ $\textmu m, a nanowire taper geometry with $w_\mathit{t}=200\nm$ was chosen to ensure the tapering nanowire effective index crossed that of the constant diameter nanofibre, as shown in Fig. \textbf{\ref{Fig.1:}}\textbf{b}. $w_\mathit{s}=700\nm$ was chosen to ensure single-mode operation of our waveguides after TOX removal. Nearly $100\%$ coupling between the nanofibre region and the standard single-mode fibre (SMF) region can be achieved \cite{Orucevic_2007, Stiebeiner_2010, Hoffman_2014, Nagai_2014}; here, we use the minimum taper geometry as described in \cite{Nagai_2014}. Any sub-unit fibre taper efficiency, $\eta_\mathit{fib}$, can be approximately compounded with the rest of the coupler transmission, $\eta_\mathit{coupler}$ to obtain the total coupler efficiency: $\eta_\mathit{tot} = \eta_\mathit{fib}+\eta_\mathit{coupler}$. The fibre taper region is terminated prior to the coupling with the silicon nanowire, so can be neglected in the adiabatic transfer simulations; the nanofibre diameter is constant over the length of the tapering silicon nanowire.

Fig. \textbf{\ref{Fig.1:}}\textbf{b} illustrates the effective indices of the coupled structure supermodes (squares, diamonds) and the fundamental modes of the separate fibre and nanowire (triangles, circles respectively) as a function of waveguide width $w$ and propagation distance $z$. Due to the resulting mode anti-crossing, the odd supermode is only supported for $w\gtrsim550\nm$. Therefore, using the even supermode ($n_\mathit{eff_\mathit{e}}\geq1.08158$) and the free-space mode ($n_\mathit{eff_\mathit{o}}=n_\mathit{fs}=1$), the corresponding beat length is calculated as $L_\mathit{b}=25.39\ $\textmu m \cite{Tiecke_2015}. $L_\mathit{c}=30\ $\textmu m was chosen for our final design to ensure $L_\mathit{c}>L_\mathit{b}$. Mode field profiles of the coupled system are shown in Fig. \textbf{\ref{Fig.1:}}\textbf{c}, illustrating the transfer of the fundamental mode from the beginning ($z=0$) of the coupling region to its end ($z=L_\mathit{c}=30\ $\textmu m). In this work we focus on TE-like modes in the nanowire, however, an equivalent coupler for TM-like modes is also possible, provided $t_\mathit{BOX}$ is large enough to prevent coupling into the silicon handle. 

\begin{figure*}[ht!]
	\centering
	\includegraphics[width = 1.0
	\linewidth]{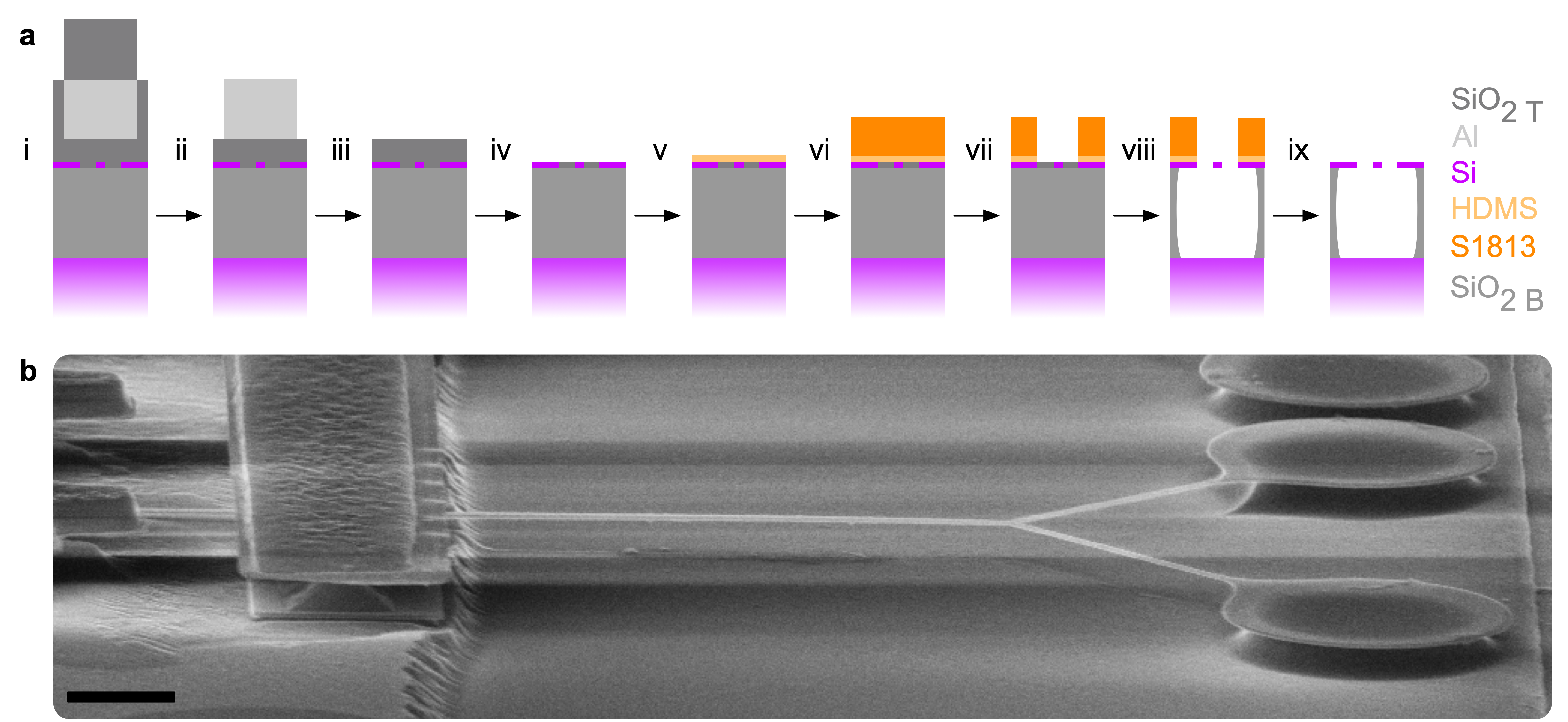}
	\caption{Foundry-fabricated photonic integrated circuit (PIC) post-process flow for suspending waveguide structures. \textbf{a} Post-process flow steps, explained in the main text. \textbf{b} Scanning electron micrograph (SEM) of final suspended silicon nanowire on the PIC. Scale bar $10\ $\textmu m.
	}
	\label{Fig.2:}
\end{figure*}

We simulate the coupler transmission using three-dimensional finite-difference time-domain simulations (Lumerical, FDTD), and display the results in Figs. \ref{Fig.1:}\textbf{b}. We plot the simulated transmission from the fundamental $\mathrm{LP_{01}}$ fibre mode into the fundamental $\mathrm{TE_0}$ nanowire mode, with $2.071\ $\textmu m light input from the fibre side, for varying waveguide width $w$, and corresponding propagation distance $z$. For a coupler without mechanical supports, the  field transmission at the end of the coupling region is above --$0.07\dB$ as calculated in 3D FDTD. However, the mechanical supports are required to ensure the silicon nanowire remains suspended, and they have a significant negative effect on the transmission. We designed supports with strut widths of $200\nm$, extending at an angle of $20^\circ$ from the fibre normal for $16\ $\textmu m until they connect with the support islands, which were $15\ $\textmu m in diameter. These support struts, whilst essential for suspension, reduce the theoretical coupler transmission to --$1.69\dB$.

%%%%%%%%%%%%%%%%%%%%%%%%%%  Devices and fabrication  %%%%%%%%%%%%%%%%%%%%%%%%%%

\section{Devices and Fabrication} \label{sec:dev_fab}

We fabricate the adiabatically tapered conic nanofibre tips using a conventional hydrogen flame-brush method described in Ref.\cite{Nagai_2014}. For our experiments, we design the optimal fibre taper profiles using the three-layer fibre model \cite{Karapetyan_2012, Nagai_2014}, and fabricate the taper from commercial germanium-doped silica single-mode fibre (SMF, SM2000 Thorlabs), which has an NA of $0.12$ and a quoted mode field diameter of $13\ $\textmu m at a wavelength of $1.996\ $\textmu m. The fabrication method (similar to Ref.\cite{Tiecke_2015}) can be divided into two parts. First, a `biconic' tapered fibre is created by heating and pulling the SMF according to the desired taper profile, leading to a waist diameter and total biconic taper length of $1.06\ $\textmu m and $14.2\mm$, respectively. Second, after removing the heat, two conic fibres are formed by pulling along the fibre axis with a motorised stage, yielding the final nanofibre taper length of about $7.1\mm$, the last $\approx200\ $\textmu m of which has a constant diameter. After pulling the nanofibres, the non-tapered end of the fibre was spliced to conventional Corning SMF-28 fibre with angled ferrule connectors for interfacing with other equipment.

The suspended tapering silicon waveguides are formed as follows. Silicon waveguides are patterned by optical lithography and dry etching at a commercial foundry (Advanced Micro Foundry, AMF Singapore). After receiving the foundry-fabricated PICs, we post-process them to suspend the tapered silicon nanowires. The post-process flow is illustrated in Fig.~\textbf{\ref{Fig.2:}}\textbf{a}, and we highlight steps i--ix inline, below. We start with a full stack PIC, with silicon handle, $t_\mathit{BOX}=3\ $\textmu m of BOX, $t_\mathit{si}=220\nm$ thick silicon device layer, $3\ $\textmu m of TOX, and a $2\ $\textmu m layer of aluminium embedded in the TOX (i). Firstly, the TOX is wet-etched with 7:1 buffered oxide etchant (BOE) down to the base of the aluminium layer ($\approx2.1\ $\textmu m, ii). To remove the aluminium, a wet etchant mixture of 16:1:1:2 phosphoric acid, nitric acid, acetic acid, and water, heated to $50^{\circ}$C is used (iii). We found a thin layer ($\approx25\nm$) of undocumented tantalum nitride was present in our foundry-fabricated PIC, for silica passivisation. This layer remained during our measurements, suspended ($\approx2\ $\textmu m) above the silicon device layer wherever residual TOX was present. Next, we remove the remaining TOX ($\approx0.9\ $\textmu m) with a further BOE wet etch (iv). Once the surface of the silicon device layer is exposed, we clean it with acetone/isopropyl-alcohol and hard bake at $150^{\circ}$C for 5 minutes. We then apply a monolayer of hexamethyldisilazane (HDMS) in vapour phase as an adhesion promoter for our photoresist (v). Next, a layer of S1813 positive photoresist, $\approx1.3\ $\textmu m thick, is spun on top of the HDMS, and soft-baked at $115^{\circ}$C for 1 minute (vi). We pattern the photoresist with a Heidelberg uPG101 maskless direct laser writer. Once patterned, the resist is developed with Microposit MF-319 photoresist developer to reveal sections of BOX for subsequent etching (vii). The PIC is then submerged in BOE again to etch the entire BOX thickness, wherever resist windows have been opened (viii). Finally, the resist and adhesion promoter are removed with an acetone/isopropyl-alcohol 2-step clean, and hard-baked at $150^{\circ}$C for 5 minutes, resulting in the final suspended structures (ix).

%%%%%%%%%%%%%%%%%%%%%%% Measurements %%%%%%%%%%%%%%%%%%%%%%%%%

\section{Measurements} \label{sec:coup_trans}

\begin{figure*}[!ht]
	\centering
	\includegraphics[width = 1.0
	\linewidth]{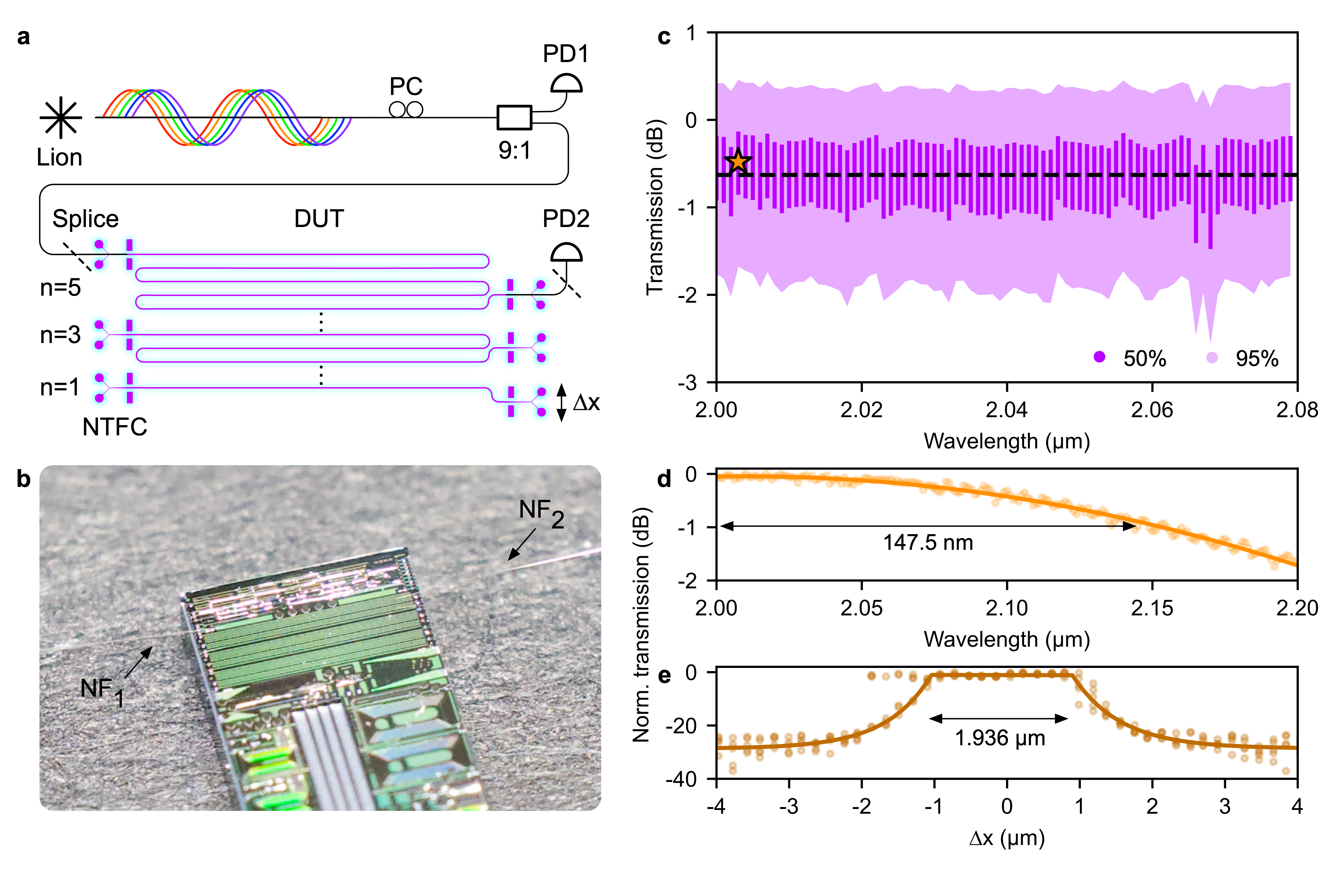}
	\caption{
	Single coupler transmission, bandwidth and lateral misalignment measurements. \textbf{a} General experimental setup. A tunable continuous-wave laser (Sacher Lasertechnik Lion) pumps an in-line fibre polarisation controller (PC), which is then tapped by a 90:10 (9:1) fibre beamsplitter. Tapped ($10\%$) power is collected by a mid-infrared (MIR) photodiode (PD1). Majority ($90\%$) of power is coupled into the device under test (DUT) and out-coupled into a final MIR photodiode (PD2). Angled intersects on DUT input/output fibres represent SMF-28 to SM2000 fusion splices, a source of loss when calculating single-coupler transmission. DUTs with $n=1, 3, 5$, labelled for clarity. Lateral displacement axis denoted by $\Delta x$. \textbf{b} Post-processed DUT with nanofibres (NF) aligned from the left and the right of the DUT for optical input/output. \textbf{c} Calculated transmission of a single nanotapered fibre coupler. Maximum calculated coupling of $-0.48\dB$ (maximum; 95\% confidence interval [0.46, --1.68]$\dB$) is indicated by a star. Dark bars indicate $50\%$ confidence interval, shaded area indicates $95\%$ confidence interval. Dotted black line indicates mean coupling across the measured wavelength range ($-0.63\dB$). \textbf{d} Normalised extended bandwidth measurement of a single nanotapered coupler. Inferred 1-$\dB$ bandwidth of $295\nm$ (3-$\dB$ bandwidth of $511\nm$). \textbf{e} Normalised transmission of the two nanotapered coupler system with varying degrees of lateral misalignment on one of the couplers. A flat response is seen for displacement values of $\pm0.968\ $\textmu m.
    }
	\label{Fig.3:}
\end{figure*}

Special care is required to characterise devices with transmission approaching unity ($0\dB$), as transmission or intensity uncertainty in the characterisation apparatus has a significant effect. To measure the single-coupler transmission, we used the cutback method. This method allows us to null the effects of on-chip propagation and bend losses, helping to isolate the coupler loss, which should not depend on waveguide length or number of bends. We utilise a 9:1 fibre beamsplitter in our experimental setup to normalise the transmission of the device under test (DUT) to the input power (Fig. \ref{Fig.3:}\textbf{a}). If we assume that the excess loss on the output arms of the 9:1 beamsplitter are identical, as is the efficiency of the two photodiodes, then the only outstanding off-chip losses are the combined losses of the nanofibre tapers and splices, $\eta_{\mathit{splice}_{1}}$ and $\eta_{\mathit{splice}_{2}}$, which were measured as $-0.07\dB$ and $-0.09\dB$ for each of the nanofibres respectively. These are subtracted from the coupler transmission values which follow. 

The cutback devices included $n=1, 3, 5$ concatenated copies of a simple unit cell, with a NTFC on the chip input and output. Each unit cell contains either two 90-degree bends or one 180-degree bend and a straight waveguide section as seen in Fig. \textbf{\ref{Fig.3:}}\textbf{a}, resulting in cutback lengths $2.71\mm$, $8.04\mm$, and $13.4\mm$. Bends were of the low-mode-conversion Euler geometry \cite{Dulkeith_2004}, parameterised following the treatment in \cite{Vogelbacher_2019}, with $p=1$. Whilst a pair of 90-degree Euler bends is morphologically distinct from a single 180-degree Euler bend, by specifying a large enough bend radius ($R_\mathit{min}=10\ $\textmu m in our structures), and setting $p=1$ to ensure identical path lengths, the transmission of these bend structures can be assumed to be indistinct. Waveguides were of the strip geometry throughout, and had a constant width of $700\nm$. For clarity, since each device had a NTFC at both the input and the output, each measured light path contained the insertion loss of two NTFC structures. Input and output coupler pairs were offset from one another to ensure that direct fibre-fibre coupling was minimised. By ensuring the number of equivalent 90-degree bends in the test structures increased linearly with the number of straight sections, $n$, as shown in Fig. \textbf{\ref{Fig.3:}}\textbf{a}, the total on-chip transmission of the cutback measurements is given by $\eta_{\mathit{chip}}(n)=2\eta_{\mathit{NTFC}}+n(\eta_{\mathit{straight}}+2\eta_{\mathit{bend}})$. Accounting for $\eta_{\mathit{splice}_{1}}$ and $\eta_{\mathit{splice}_{2}}$, along with all uncertainties in power splitting and detection off-chip (dominated by the $\pm5\%$ photodiode measurement uncertainty), single-NTFC transmission with low-uncertainty was measured. 

Our experimental setup is shown in Fig. \textbf{\ref{Fig.3:}}\textbf{a}. We used a tunable continuous-wave (CW) laser (Sacher Lasertechnik Lion) and integrating sphere power meter (Thorlabs S148C) to obtain transmission spectra. The laser was fibre coupled into fibre, polarisation controlled (FiberPro PC1100) and then tapped off with a 90:10 fibre beamsplitter (Thorlabs TW2000R5A1B). Light from the $10\%$ tap was collected with a MIR photodiode within an integrating sphere (PD, Thorlabs S148C) whilst light from the $90\%$ arm was coupled via an NTFC into the device under test (DUT). Finally the output of the DUT was coupled back into fibre via a second NTFC and collected with a second, identical MIR PD.

Both input and output nanotapered fibres were mounted to 6-axis translation stages (Thorlabs Nanomax MAX607) on rotating fibre mounts (Thorlabs HFR007). The mounting stages were pitched forward, with tips $3^{\circ}$ downward, as this was found to have a positive impact on the transmission of our devices. Each of the 12 degrees of freedom on the translation stages were computer-controlled with a precision stepper motor actuators (Qontrol ACTL2). An image of a post-processed PIC with input NTFCs can be seen in Fig. \textbf{\ref{Fig.3:}}\textbf{b}.

By measuring the transmission at each wavelength as a function of cutback length, then fitting a linear trend, we were able to extract the $n=0$ transmission spectrum. By subsequently subtracting the splice losses ($\eta_{\mathit{splice}_{1}}$, $\eta_{\mathit{splice}_{2}}$) and dividing by the number of couplers in the device (2), we could estimate the transmission of a single NTFC, as shown in Fig. \textbf{\ref{Fig.3:}}\textbf{c}. Here, we denote 50\% and 95\% confidence intervals as bars and shading respectively. We observe a maximum coupler efficiency of $\text{--}0.48\ \substack{+0.94\ \\\text{--}1.2\ }\dB$ (log-normal distribution), and a mean coupler transmission of --$0.63\dB$ across the range 2.00--2.08$\ $\textmu m. 

In Fig. \textbf{\ref{Fig.3:}}\textbf{d}, an extended spectrum was measured with the aid of an additional longer-wavelength tunable CW laser (Sacher Lasertechnik Lion), resulting in a fitted $295\nm$ 1-$\dB$ bandwidth ($511\nm$ 3-$\dB$ bandwidth), normalised to peak transmission (second-order polynomial fit). Having measured insertion loss and bandwidth of our NTFCs, we finally proceeded to measure its lateral misalignment tolerance. For this, we fixed the probe wavelength at $2.071\ $\textmu m and coupled into the $8.04\mm$ cutback device. Once coupled, an autonomous coupling-uncoupling routine was executed 250 times, with varying amounts of lateral displacement ($\Delta x$) in each iteration. Results of this investigation (normalised to peak transmission) can be seen in Fig. \textbf{\ref{Fig.3:}}\textbf{e}. A flat response can be seen for displacement values of $\pm0.968\ $\textmu m, suggesting good lateral misalignment tolerance of the coupler.  

%%%%%%%%%%%%%%%%%%%%%%% End of body %%%%%%%%%%%%%%%%%%%%%%%%%

%%%%%%%%%%%%%%%%%%%%%%% Discussion %%%%%%%%%%%%%%%%%%%%%%%%%

\section{Discussion} \label{sec:disc}

We have demonstrated couplers which exhibit an efficiency of $-0.48\ \substack{+0.94\\-1.2\ }\dB$, a $295\nm$ 1-dB bandwidth, and a $\pm0.968\ $\textmu m misalignment tolerance. Our couplers occupy a small footprint, and require only simple back-end wet-etches on a foundry-fabricated PIC.

Theoretically, the coupler transmission of our devices should have been a relatively poor $-1.69\dB$, whereas in practice we measured the much better $-0.48\dB$. One explanation is that the initial simulations of coupler efficiency did not account for the pitch of the nanofibre tip with respect to the PIC surface: in experiment, the nanofibre tip was pitched at an angle of $-3^{\circ}$, which we found had a positive impact on the transmission. We theorise that this pitch reduced the interaction of the guided mode with the support structures, enhancing coupler transmission. It may be useful to design future coupling systems to include this pitch.

We found that support structures were essential to the mechanical suspension of the on-chip tapering nanowires. In future, the optical design of these could be improved. Future devices could be suspended via partial etches. Shallow or deep partial etches could be employed, in conjunction with subwavelength gratings, to laterally support the nanowire. This would ensure minimal interaction with the guided mode, whilst providing a more natural solution for suspended nanowire yield and integrity \cite{Nedeljkovic_2020}. Another solution, more similar to that utilised here, could involve more careful consideration of the support geometry. By tethering perpendicular to the direction of propagation, and appropriately selecting the tether waveguide width, an anti-reflection $\lambda/4$ layer could reduce reflections introduced in the coupling system, at the cost of device bandwidth.

We initially chose to include a metal layer above the NTFC dams to act as an alignment marker for resist patterning with electron-beam lithography. However, this was found to be unnecessary for the direct laser writer, which was far quicker and more convenient. The inclusion of non-essential layers in the proximity of the device should be avoided.

\begin{table}[ht]
    \centering
    \caption{NTFC design parameters for $\lambda=2.071\ $\textmu m on the silicon nitride (SiN) and lithium niobate (LiNbO$_{3}$) on insulator material platforms for standard film thicknesses, $t_\mathit{film}$. $L_\mathit{c}>L_\mathit{b}$, all values in \textmu m. Assumes top and bottom oxide removal. Bottom oxide thickness $3\ $\textmu m.}
    \begin{tabular}[t]{lccccc}
        \hline
        Material &\ $t_\mathit{film}$ &\ $w_\mathit{t}$ &\ $w_\mathit{s}$ &\ $L_\mathit{b}$ \\
        \hline
        $\mathrm{SiN}$ & 0.3 & 0.4 & 1.2 & 22.08 \\
        $\mathrm{LiNbO_{3}}$ & 0.6 & 0.3 & 1.2 & 22.56 \\ 
        \hline
        \label{multi_material_params}
    \end{tabular}
\end{table}

So long as the nanofibre and waveguide mode share an effective index crossing as shown in Fig. \textbf{\ref{Fig.1:}}\textbf{b}, our design can be readily extended to any high-index contrast platform. For example, high performance photonic interconnects between silicon, low-loss silicon nitride, and $\chi^{(\mathit{2})}$ lithium niobate could be achieved for photonics applications that require multi-material functionality. NTFC parameters for thin-film lithium niobate and silicon nitride platforms can be found in Tab. \textbf{\ref{multi_material_params}}.\\
\indent The work conducted here is the first example of low-loss, broadband fibre-chip coupling on a high-index contrast platform in the MIR. The methodology employed in this investigation can readily be applied to different material platforms and wavelengths of operation. With this work we pave the way to exciting applications on SOI in the MIR, and more generally, for modular integrated photonic systems where different material platforms can be interfaced with low-loss fibre interconnects to make the most each platform's unique strengths .

%%%%%%%%%%%%%%%%%%%%%%% Acknowledgements %%%%%%%%%%%%%%%%%%%%%%%%%

\section*{Acknowledgements} \label{sec:acknowledgements}
% \vspace{-1mm}
    We would like to thank the invaluable insight and advice of Ankur Khurana, Quinn M. B. Palmer and Vinita Mittal in the post-processing of the devices used in this work. Also, thanks to Jonathan Frazer for fruitful discussions regarding error analysis, to Jorge Monroy Ruz for providing silicon nitride dispersion data, and to the AMF team, and to Bu Xiao Mei in particular. The authors gratefully acknowledge the following grant funding: Leverhulme Trust (ECF-2018-276), UK Research and Innovation Future Leaders Fellowship (MR/T041773/1), QuPIC (EP/N015126/1). Y.Y. acknowledges the Leading Graduate Program in Science and Engineering, Waseda University (A12621600) awarded by the Ministry of Education, Culture, Sports, Science and Technology (MEXT), Japan. T.A. acknowledges the support of the grants: Japan Science and Technology CREST (JPMJCR1771), and Japan Society for the Promotion of Science KAKENHI (18H05207).

%%%%%%%%%%%%%%%%%%%%%%% References %%%%%%%%%%%%%%%%%%%%%%%%%

\bibliography{bib}

%%%%%%%%%%%%%%%%%%%%%%% End of doc %%%%%%%%%%%%%%%%%%%%%%%%%

\end{document}